\documentstyle[12pt,aasms4]{article}

\def\msun{{\rm\,M_\odot}}

\def\vol#1  {{{#1}{\rm,}\ }}
\newcommand{\etal}{et~al.~}

\begin{document}

\title{Towards Understanding Galaxy Clusters and Their Constituents:
Projection Effects on Velocity Dispersion, X-Ray Emission,
Mass Estimates, Gas Fraction and Substructure}
\author{Renyue Cen\altaffilmark{1,2}}

\centerline{cen@astro.princeton.edu}


\altaffiltext{1} {Princeton University Observatory, Princeton University, Princeton, NJ 08544}
\altaffiltext{2} {Department of Astronomy, University of Washington, Seattle, WA 98195}

\begin{abstract} 
We study the projection effects on various observables of clusters
of galaxies at redshift near zero,
including cluster richness, velocity dispersion,
X-ray luminosity, three total mass estimates
(velocity-based, temperature-based
and gravitational lensing derived),
gas fraction and substructure,
utilizing a large simulation of a realistic cosmological model
(a cold dark matter model with the following parameters: 
$H_0=65$km/s/Mpc, $\Omega_0=0.4$, $\Lambda_0=0.6$, $\sigma_8=0.79$).
Unlike previous studies focusing on the Abell clusters,
we conservatively assume that both optical and X-ray observations
can determine the source (galaxy or hot X-ray gas)
positions along the line of sight as well as in the sky plane accurately; 
hence we only include sources
inside the velocity space defined by the cluster galaxies (filtered
through the pessimistic $3\sigma$ clipping algorithm)
as possible contamination sources.
Projection effects are found to be important for some
quantities but insignificant for others.

We show that, on average, the gas to total mass ratio
in clusters 
appears to be 
30-40\% higher than its corresponding global ratio.
Independent of its mean value,
the broadness of the observed distribution
of gas to total mass ratio 
is adequately accounted for by projection effects,
alleviating the need to invoke
(though not preventing) other
non gravitational physical processes.
While the moderate boost in the ratio
narrows the gap,
it is still not quite sufficient to 
reconcile the standard nucleosynthesis value
of $\Omega_b=0.0125(H_0/100)^{-2}$ (Walker \etal 1991) and $\Omega_0=1$ 
with the observed gas to mass ratio value in
clusters of galaxies, $0.05(H_0/100)^{-3/2}$, 
for any plausible value of $H_0$. 
However, it is worth noting that real observations of X-ray clusters,
especially X-ray imaging observations,
may be subject to more projection contaminations than we 
allow for in our analysis.
In contrast, the X-ray luminosity of a cluster 
within a radius $\leq 1.0h^{-1}$Mpc
is hardly altered by projection,
rendering the cluster X-ray luminosity function 
a very useful and simple diagnostic for comparing observations
with theoretical predictions.

Rich cluster masses [$M(<1.0h^{-1}Mpc)\ge 3\times 10^{14}h^{-1}\msun$]
derived from
X-ray temperatures or galaxy velocity dispersions 
underestimate, on average,
the true cluster masses by about 20\%,
with the former displaying a smaller scatter, 
thus providing a better means for cluster mass determination.
The gravitational 
lensing reconstructed (assuming an ideal inversion) mass 
is, on average,
overestimates the true mass by only 5-10\%
but displays a dispersion significantly larger than
that of the X-ray determined mass. 
The ratio of the lensing derived mass to 
the velocity or temperature derived mass
is about 1.2-1.3 for rich clusters,
with a small fraction reaching about $\sim 2.0$.
The dispersion in that ratio increases rapidly for poor clusters,
reaching about 1.0-2.0 for clusters with masses 
of $M\sim 1-3\times 10^{14}\msun$.
It appears that projection effects alone may be able
to account for the disparities 
in existing observational data for cluster masses, 
determined by various methods.

Projection inflates substructure measurements in galaxy maps,
but affects X-ray maps much less.
Most clusters ($\ge 90\%$) in this model universe
do not contain significant intrinsic substructure 
on scales $\ge 50h^{-1}$kpc at $R_{proj} \le 1h^{-1}$Mpc
without projection effects,
whereas more than $\sim 50\%$ of the same
clusters would be ``observed" to show
statistically significant substructure as measured 
by the Dressler-Shectman $\Delta$ statistic.
The fact that a comparable fraction ($\sim 50\%$)
of real observed clusters show substructure measured in the same way
implies
that most of the substructure
observed in real clusters of galaxies may be due to projection.

Finally, we point out that it is often very 
difficult to correctly interpret complex structures
seen in galaxy and X-ray maps of clusters,
which frequently display illusory configurations due to projection.
Until we can determine real distances of X-ray sources
and galaxies accurately,
for some observables, 
the only meaningful way to compare predictions of 
a cosmological model with the cluster observations 
is to subject clusters in a simulated universe 
to exactly the same observational biases and uncertainties,
including projection and other instrumental limitations,
and to compare the ``observed" simulated clusters
with real ones.

\end{abstract}


\section{Introduction} 

Among the most striking features 
found in large-scale
galaxy redshift surveys are the filaments and walls
of galaxies, stretching up to $\sim 100$
megaparsecs (Geller \& Huchra 1989) and surrounding large
under populated semi-spherical regions.
X-ray observations also hint that hot ($T\ge 10^6$~Kelvin)
diffuse gas occupies much larger intracluster 
and intercluster space 
than the isolated cores of clusters of galaxies (Soltan \etal 1996).
Numerical simulations of growth of large-scale structure 
within the gravitational instability paradigm
have produced a remarkably successful account of 
the large-scale structure observations,
reproducing large-scale structural features such as the ``Great Wall" 
of galaxies (Park 1990; Cen \& Ostriker 1993)
as well as clusters of galaxies, which most often
reside at the intersections of filaments and walls 
(Kang \etal 1994; Bryan \etal 1994; Cen \& Ostriker 1994). 
It follows, then, that clusters of galaxies should display 
different properties when viewed at different angles.
For instance,
a cluster of galaxies, embedded in a wall,
is likely to contain substantially more projected,
external structures
when the line of sight lies
in the wall than when the line of sight 
is perpendicular to the wall.
As a consequence, it is more difficult in the former than 
in the latter 
to correctly identify/interpret
genuine cluster member galaxies, intrinsic intracluster gas,
genuine substructure, depth of the cluster potential well, etc.

This is not the first time that projection
effects have been suggested.
In fact, it has long been suspected that 
systems such as the Abell galaxy clusters (Abell 1958), 
which are identified as large enhancements in the surface 
number density of galaxies on the sky,
are substantially contaminated (Lucey 1983; Frenk \etal 1990).
On a somewhat smaller scale, Hickson Compact Groups (HCGs; Hickson 1982)
have received considerable attention.
Rose (1977) first suggests that 
chain-like HCGs may not be bound,
and subsequently it is argued that HCGs could also be
chance (unbound) 3D configurations within loose groups (Rose 1979). 
Mamon (1986) goes further to claim
that most HCGs are not bound.
Recently, Hernquist, Katz, \& Weinberg (1995)
suggest that many (and perhaps most) of 
HCGs are projected configurations along long filaments.

In this paper we study the effects of
projection on gravitationally bound galaxy clusters 
by foreground and background objects,
utilizing a large-scale numerical
simulation of a realistic cosmological model.
The adopted model is a cold dark matter model with 
a cosmological constant,
which has been shown to be consistent with
most galaxy and large-scale observations in our local universe
(Cen, Gnedin, \& Ostriker 1993; Ostriker \& Steinhardt 1995) as
well as with high redshift Lyman alpha forest observations
(Cen \etal 1994; Miralda-Escud\'e 1996; Rauch \etal 1997).
We wish to examine the projection effects 
on several properties of clusters of galaxies
observed either in galaxies or in X-ray gas.
In the former case with galaxies,
we assume that each galaxy 
(above a certain magnitude limit)
is measured accurately both in the sky plane 
and in redshift (velocity) space,
in sharp contrast with previous studies
of Abell clusters where no redshift information is used.
In the latter case with X-ray gas,
X-ray observations of clusters of galaxies by satellite telescopes 
such as  ROSAT (Tr$\ddot u$mper 1990) or ASCA (Tanaka \etal 1994)
with limited spectroscopic capabilities
may have severely confused foreground and background X-ray luminous
objects with the intrinsic structures 
physically within clusters themselves.
However, we take a somewhat conservative approach to this problem. 
While X-ray observations of clusters 
of galaxies have lower velocity resolution,
thus
allowing for more distant objects along the line of sight to be projected,
we assume (rather generously perhaps)
that optical follow-up observations would help
remove all distant X-ray projection 
beyond the velocity domain defined by the cluster galaxies.

This paper is organized as follows.
In the next section we give a brief description
of the simulation and a detailed description of 
a method to generate mock cluster catalogs.
Results are presented in \S 3, where we first present
qualitative pictures, then statistical analyses covering
five topics in separate subsections.
In \S 3.2.1 we describe 
the projection effects on galaxy cluster richness. 
In \S 3.2.2 the projection effects 
on X-ray luminosity are presented.
In \S 3.2.3 the projection effects on
velocity dispersion and various mass estimates
are discussed.
In \S 3.2.4 we examine projection effects on 
gas to total mass ratio.
In \S 3.2.5 we examine how projection effects
affect the interpretation of substructure.
Discussion and conclusions are 
given in \S 4.

\section{Generating Mock Cluster Catalogs}

\subsection{N-body Simulation}
A P$^3$M simulation made on a GRAPE board based Sun Sparc 10
workstation is used (Brieu, Summers \& Ostriker 1995).
We simulate a cold dark matter universe with
the following parameters (Summers, Ostriker, \& Cen 1997):
$H_0=65h^{-1}$Mpc,
$\Omega_0=0.4$,
$\Lambda_0=0.6$,
$\sigma_8=0.79$
(with a slight tilt of the spectral index, $n\sim 0.95$,
and a gravity wave contribution of 25\%).
The box size is $128h^{-1}$Mpc with $256^3$ cells,
$128^3$ particles and a Plummer softening length
of 0.05 cell, giving a true spatial resolution ($\sim 2.0$ Plummer
softening lengths)
of $50h^{-1}$kpc and a mass resolution of $1.1\times 10^{11}h^{-1}\msun$.
The power spectrum transfer function is computed
using the method described in Cen, Gnedin, \& Ostriker (1993).
Gaussian initial conditions are used.
The simulation starts at $z=40$
and the simulation box at $z=0$ is used for the analysis below.

The primary motivation for choosing this model
is that it best fits the available observations.
Specifically, it is consistent with the following observations
(Ostriker \& Steinhardt 1995).
At high redshift, the model is consistent
with the current leading constraint
from COBE observations (\cite{s92}),
which fixes the amplitude
of the power spectrum on very large scales ($\sim 1000h^{-1}$Mpc)
to an accuracy of about 12\%.
In particular, the model is normalized to the first year COBE observations
(Kofman, Gnedin, \& Bahcall 1993).
At low (essentially zero) redshift, we demand that 
the model fits 
current observations of our local 
universe, primarily those concerning 
the distributions of galaxies and hot X-ray emitting gas
in $(\vec x, \vec v)$ space,
including 1) the abundance of clusters of galaxies, 
which fixes the amplitude
of the power spectrum on scales of $\sim 8h^{-1}$Mpc
to about 10\% accuracy (Bahcall \& Cen 1992,1993;
\cite{ob92};
\cite{wef93};
\cite{co94};
\cite{vl95};
\cite{bm96};
\cite{ecf96};
Pen 1996),
2) the power spectrum of galaxies, which constrains the shape 
of the power spectrum on the intermediate-to-large scales,
$\sim 10-100h^{-1}$Mpc (Peacock \& Dodds 1994; Feldman, Kaiser \& Peacock 1994),
3) the ratio of gas to total mass in galaxy clusters, 
which constrains $\Omega_b/\Omega_{tot}$ (\cite{wnef93}; \cite{lcbo96}).
In addition, the model is consistent with
the current measurements of the Hubble constant 
(Fukugita, Hogan, \& Peebles 1993;
Freedman \etal 1994;
\cite{rpk95};
Hamuy \etal 1995)
and the age constraint from the oldest globular clusters 
(Bolte \& Hogan 1995) as well
as the gravitational lensing constraint
(\cite{ft91}; \cite{mr93}; \cite{cgot94}; \cite{k96}).

\subsection{Selecting Clusters of Galaxies}

Particles in the simulation box (cube) at $z=0$ are
projected along each of the three orthogonal axes.
Then, high concentrations of particles 
within circles of radius $1h^{-1}$Mpc
in the projected, two-dimensional distributions 
are rank-ordered by the number of particles inside each circle.
The top $50$ densest two dimensional clusters 
(they are columns of depth $128 h^{-1}$Mpc)
along each of the three
projections (a total of 150 clusters)
are selected for further examination.
So far, we have selected 2-d clusters ignoring any 
redshift information of a particle, a procedure
similar to what is used to select Abell clusters.
Each column has an area of size $6\times 6h^{-2}$Mpc$^2$
centered on the 2-d center of the projected particle
distribution,
a size large enough to allow possible offset of the 
center of the main cluster in the column.
For each particle contained in these columns
the following information is registered:
sky plane position ($x,y$), line-of-sight distance $r$ relative
to the midplane of each column, 
radial velocity $v_r$, 
mass density $\rho$ 
[obtained using SPH smoothing kernel with 64 neighbor particles;
Katz (1996, private communication)
and Stadel (1996, private communication]
and local velocity dispersion $\sigma$
(obtained in a similar way as for $\rho$).

Next, we find the line-of-sight position of the largest cluster
in each column, and place it at a distance of $100h^{-1}$Mpc from us.
The line-of-sight positions of all the particles in the column
are then re-arranged to center on the main cluster in the column,
which is easily accommodated since the simulation box is periodic.
The sky plane position of the center of 
the main cluster is calculated and the column is re-centered
on the cluster; in most cases the offset is small.

In order to make direct comparisons with observations,
we generate mock galaxy cluster catalogs by assigning
each particle a luminosity drawn randomly from
a standard Schechter luminosity function (Schechter 1976);
i.e., N-body points are assumed to be unbiased.
We take the values of
$\phi_*=0.0275$, $\alpha=-1.1$, $M_*=-19.2$,
obtained by averaging over the CfA redshift survey 
(Marzke, Huchra, \& Geller 1994)
and the southern sky redshift survey (SSRS2; Da Costa \etal 1994).
Note that for this purpose we consistently use
$h=1$ by adopting a length unit of $h^{-1}$Mpc.
When we examine the properties of clusters of galaxies
in such a mock catalog, only galaxies 
with apparent magnitudes of 18.0 or less 
in each cluster (at a distance of $100h^{-1}$Mpc)
are included for analysis, mimicing the spectroscopic apparent magnitude limit
of the Sloan Digital Sky Survey (\cite{gk93}; \cite{gw95}). 
Note that placement of the clusters at $100h^{-1}$Mpc is 
rather arbitrary and 
merely for the convenience of illustration and explanation.

Furthermore,
for each cluster we apply the pessimistic $3\sigma$ clipping
scheme (Yahil \& Vidal 1977) to remove outliers.
Here we briefly describe the
pessimistic $3\sigma$ clipping method.
Galaxies within a projected radius $R_{proj}$ about the center of a cluster 
are rank-ordered 
according to their distances
from the center of the cluster in velocity space.
Then the most distant galaxy 
from the cluster center in velocity space 
is removed, {\it if it is more distant than
$3\sigma$ from the cluster center},
where $\sigma$ is the velocity dispersion of the cluster
in the cluster center frame
computed by
excluding that most distant galaxy.
This procedure is repeated until no more galaxies can be removed.
The remaining galaxies are then 
``observed" as members of the cluster and analysed.
Of course, we wish to see the projection effects, 
so the ``true" members of each cluster are 
separately identified
with those belonging to the cluster
when the DENMAX grouping scheme (\cite{bg91})
is used in real 3-dimensional space.
A Gaussian smoothing of radius $250h^{-1}$kpc is applied before
the DENMAX algorithm operates.
Briefly, the DENMAX scheme works as follows.
Each particle is moved along the gradient of the
density field (defined on a grid of $512^3$ points in this case)
until it reaches a local density maximum.
Particles collected at separate local density maxima
are grouped into separate objects, which
we call real clusters.

Before leaving this section, a few words concerning the
limitations of our simulated galaxy cluster catalogs
are appropriate.
Due to the limited mass resolution 
($1.1\times 10^{11}h^{-1}\msun$),
sub-$L^*$ galaxies/halos are not resolved in the
current simulation.
For this reason, we have chosen to select ``galaxies"
from the simulation by randomly sampling the particle distribution,
rather than directly selecting out halos, which would be
more physically motivated were all the relevant halos properly resolved.
The hope is that the velocity field in and around
clusters is adequately sampled this way, and insofar as galaxy density
bias is not a strong function of local density 
for concerned regions of sufficiently high density in this study,
the galaxy density distribution 
should be reasonably represented.

\subsection{Generating Cluster X-ray Maps}

Since each particle has
a local density and velocity dispersion,
as defined in
the preceding section, we can, albeit in a crude fashion,
generate X-ray emission for each particle, assuming
that gas follows dark matter and 
that gas temperature is equal to local velocity dispersion squared 
($T=\beta \mu m_p \sigma^2/k=\mu m_p \sigma^2/k$,
where $\mu=0.60$ is the molecular weight and $\beta=1$ is assumed;
see Edge \& Stewart 1991; Lubin \& Bahcall 1993).
A global gas density of $\Omega_b=0.0355$ is used,
consistent with light element nucleosynthesis (Walker \etal 1991)
for the adopted Hubble constant $h=0.65$.
Both line emission
and bremsstrahlung emission are included (Cen \etal 1994;
a code based on the work of Raymond \& Smith 1977),
assuming 1) a metallicity of $0.35$ in solar units, 
as observed in the great clusters (Edge \& Stewart 1991;
Arnaud \etal 1992,1994) and, 2)
the gas is in ionization equilibrium and optically thin
to X-ray photons.
X-ray maps are generated in two different
bands: $0.4-2.4$keV and $0.5-10.0$keV,
with the former matching the ROSAT harder band
and the latter mimicing the Einstein and ASCA bands. 
Emissivity-weighted temperature maps are also generated.
A pixel size of $50\times 50 h^{-2}$kpc$^2$ is used and
final maps are smoothed with a Gaussian window of 
radius $100h^{-1}$kpc.

In order to show the projection effects two kinds of X-ray maps
are generated: 1) X-ray maps due to gas in the clusters alone and, 
2) X-ray maps including foreground and background objects.
For the second case, only foreground and background objects
within the line-of-sight velocity domain of the cluster, 
delimited by the two most distant
galaxies on two sides of the cluster (in velocity space)
after the $3\sigma$ clipping algorithm is applied,
are included.
This is equivalent to assuming that X-ray observations 
have the same velocity resolution as optical observations,
which is, of course, untrue for available X-ray observations.
For example,
the spectroscopic resolution of ROSAT of $E/\Delta E\approx 2.5$
(corresponding to a velocity resolution of $\sim 60,000$km/s, if 
proper spectral lines could be identified),
and the lack of strong spectral features for hot luminous
clusters ($kT>2$keV) in its band, severely limits its ability to 
tell foreground and background objects.
On the other hand, ASCA has a much better
velocity resolution than ROSAT
due to its wider/higher energy coverage (allowing strong
spectral features to exist in its band, for example, various
iron lines; Bautz \etal 1994; Fukazawa \etal 1994)
and high $E/\Delta E\approx 50$ 
(corresponding to a velocity resolution of $\sim 3,000$km/s).
But even for the case of ASCA X-ray observations,
our approach is conservative, since it still has a lower
spectroscopic capability than that of optical observations.
In practice, our adopted assumption for the X-ray observations
is equivalent to assuming that optical follow-up observations
will be made for each X-ray observed cluster, which would
help remove foreground and background X-ray contaminations
outside the velocity space defined by the cluster galaxies.

We take one more conservative step
in generating the X-ray maps:
we exclude particles whose local densities
(obtained using 
the SPH smoothing kernel with 64 neighbor particles)
are below the global mean.
This step ensures that poorly defined velocity dispersions
(hence temperatures)
for those particles in the low density regions
do not add spurious effects.
Since all those particles are outside clusters in real
space along the line of sight,
within the projection radii we are investigating, 
we therefore
relatively underestimate the amount of projected structures,
if the excluded structures were X-ray luminous.

Although we analyze conservatively the simulated results
in the sense that the real projection effects would 
be larger if realistic observational situations  are carefully considered
and better, gasdynamic simulations are used,
we need to caution the reader
that the X-ray properties of the clusters 
may be significantly altered were gasdynamic effects included.

\section{Results} 

\subsection{Pictures --- Qualitative Results} 

Before turning to quantitative analyses,
let us first visually examine the galaxy clusters and their
X-ray maps. 
In order not to mislead the reader due to subjective
selection criteria, only 10
{\it randomly} selected clusters are shown. 
These clusters are selected out of the top 50 clusters identified
along the x-axis projection.
However, one human filtering (by eyeballing) is applied to 
avoid repeating nearly identical configurations.

The 10 clusters are shown in Figures (1) to (10) 
within a projection radius $R_{proj}=2.0h^{-1}$Mpc.
Each of Figures (1-10) consists of 20 Panels as follows.
Panel (1) shows the projected distribution of galaxies
in the cluster, as it would be observed, after applying the 
pessimistic $3\sigma$ clipping method (Yahil \& Vidal 1977),
whereas Panel (2) shows the ``true" member galaxies of the cluster (see \S 2.2).
Three symbols are used in Panel (1): 
solid dots for ``true" member galaxies [which is also used in Panel (2)
where only ``true" members are shown],
open circles for background galaxies,
and stars for foreground galaxies.
Panel (3) shows the X-ray surface brightness in the 0.4-2.4keV band
including all X-ray sources within the
velocity domain of the cluster, defined by the cluster galaxies.
Panel (4) shows ``true"
X-ray surface brightness of the cluster due to 
the hot intracluster gas {\it in the cluster only} (i.e., excluding
all possible foreground and background sources).
The contour levels in Panels (3,4) are
$10^{-8,-7,-6, ...}$erg/cm$^2$/sec/sr.
Panels (5) and (6) show the corresponding (emissivity-weighted)
temperature maps for Panels (3) and (4), respectively, 
with contour levels for thick curves of
$10^{7.00,7.25,7.50,7.75, ...}$Kelvin
and thin curves of $10^{6.75,6.50,6.25,6.00, ...}$Kelvin.
Panels (7) and (8) show the galaxies in real space and
in velocity space, respectively.
Panels (9,10) are similar to Panels (1,2)
but only for galaxies projected inside the virial radius of the cluster,
where the virial radius of each cluster is defined
as the three-dimensional radius within which 
the mean density of the cluster is 200 times 
the critical density of the universe.
Panels (11,12,13,14) are similar 
to Panels (3,4,5,6) but for ASCA $0.5-10.0$keV band.
Panels (15,16) are similar 
to Panels (7,8) but
only for galaxies projected inside the virial radius of the cluster.
Panel (17) shows the galaxy density distribution
in velocity space for ``true" members (solid
histogram) and projected members (dotted histogram) within
$R_{proj}=2.0h^{-1}$Mpc,
and Panel (18) shows the corresponding distributions in real space.
Panel (19) shows 
the line-of-sight velocity as a function of
the projection distance of each galaxy  relative to the center
of the cluster
[the symbols have the same meanings as in Panel (1)],
and Panel (20) shows 
the line-of-sight real space position as a function of
the projection distance of each galaxy.

Clearly, a variety of projection patterns exist.
A few general points are worth noting.

First, in almost all cases, 
cleanly or fairly cleanly separated structures
in real space, seen in Panels (7,15),
denoted as ``true" member galaxies (solid dots), background (open circles)
and foreground (stars) galaxies,
are completely mixed in velocity space [Panels (8,16)].
Two distinct physical processes interplay here.
First, the internal velocity dispersion in the cluster
disperses galaxies along the line-of-sight, producing the
familiar feature known as a ``finger-of-god".
Second, the relative infall motions of foreground/background
objects towards the main cluster, convolved with the
dispersion effect due to their own internal velocity dispersions, 
move these physically
unassociated components into (or close to)
the velocity domain
spanned by the genuine cluster galaxies, 
often completely disguising the real
space displacements [Panel (11)].
More massive clusters have larger velocity dispersions and
larger infall velocities (at fixed radii),
resulting in larger surrounding regions being affected.

Second, the projected components change the velocity
distributions of observed clusters in complicated ways.
The intrinsic velocity distribution of a cluster, which
itself is often non-Gaussian (even if one corrects for the
$3\sigma$ clipping effect),
can be {\it broadened or narrowed} by projection.
In other words, the velocity dispersion of a cluster
can increase or decrease  due to projection.
Two simple examples shall elaborate this point.
In the cluster shown in Figure 1 the 
original velocity distribution [solid histogram in Panel (17)]
is narrowed,
simply because the two infalling structures, 
whose own internal velocity dispersions 
are lower than that of the main cluster,
happen to have infall (proper peculiar) velocities that approximately cancel
Hubble expansion relative to the main cluster.
The net result is that the total (observed) velocity
distribution is composed of three overlapping velocity structures
and is narrower than 
that of the intrinsic cluster galaxies.
An opposite example is shown in Figure 10,
where the infall motion of
a rather distant structure at $\sim 13h^{-1}$Mpc 
brings it to the edge 
[$v\sim 1000$km/s; dotted histogram in Panel (11)]
of the intrinsic velocity distribution
of the cluster [solid histogram in Panel (11)],
rendering the total velocity distribution broader than
that of either of the two physically separate structures.

Finally, we note that, while the substructures in galaxy
distributions are often not easily recognizable 
by the eye due to the small number of galaxies involved,
X-ray maps are more revealing in this respect.
However, it seems to require extra care to correctly interpret 
the substructure seen in X-ray maps,
because many of these substructures are projected ones,
which often have nothing to do
with the intrinsic dynamics of the cluster.
A few concrete examples can be useful here.
In the cluster shown in Figure 1,
the apparent binary core structure seen in
both X-ray surface brightness maps [Panels (3,11)],
is caused by the projection of a background structure at a
distance of $2-5h^{-1}$Mpc from the main cluster
[open circles in Panel (7), also Panels (1) and (8)].
The main cluster has intrinsically 
a single core shown in Panels (4,12).
The velocity effect has entirely
erased the distinct identities
of the two structures  (the main cluster and the substructure)
in velocity space, as seen in Panels (8,16).
The temperature contours [Panels (5,6,13,14)]
do not exactly coincide
with the surface brightness maps and do not seem to
provide a better means to identify the sub-clumps.
Note that, while the real separation of
the two components is about $2-5h^{-1}$Mpc,
they appear to be separated by 
$\sim 0.5h^{-1}$Mpc only in projection. 
A naive
interpretation of 
the existence of such a binary structure would imply
that the main cluster is young, unrelaxed and undergoing 
a major merger.
Such an explanation is obviously 
incorrect in this case; it is true that
other structures are in the process of 
progressively merging onto the main cluster,
but they are at much larger distances than $0.5h^{-1}$Mpc.
Now let us look at the cluster shown in Figure 6.
In this case an originally smooth single structure [Panels (4,12)]
has additional 5 substructures seen in the projected X-ray maps 
[Panels (3,11)],
scattered in regions about $1-2h^{-1}$Mpc in projection from the cluster
center, due to various background/foreground objects
at distances ranging from $\sim 2h^{-1}$Mpc to $17h^{-1}$Mpc.
Lastly, we examine the cluster shown in Figure 9.
Here, a background 
structure at a distance $\sim 5h^{-1}$Mpc away [Panel (7)]
from the main cluster appears to be in the process merging
onto the cluster core from lower-left at a projected
distance of about $1h^{-1}$Mpc.
This illusory appearance, again, could be very misleading.

Our X-ray temperature maps are too crude  due to the lack of gasdynamics
in the simulation to allow strong statements to be made.
Although real temperature maps of gas in the
observed X-ray clusters (Arnaud \etal 1994)
contain a lot of
information, which may help decrypt the complex
patterns in the surface brightness maps 
(especically enabling substructure to be more easily recognizable),
they do not provide a better means to
distinguish, for example, between a cool clump in the cluster 
and a cool clump projected onto the cluster.

We see that even with our rather conservative approach to 
(perhaps overly) optimistically limit projection effects
on the X-ray maps, intriguingly misleading situations are common.
It indicates that imaging X-ray observations alone
or low capability spectral imaging observations alone (e.g.,
\cite{fj90}; \cite{dm93};
B$\ddot o$hringer \etal 1992) are especially vulnerable
to being misinterpreted.
We therefore caution that some substructure often characteristic
of apparent merging events in real observed clusters
(e.g., \cite{b91}; \cite{dm93}; \cite{m93}; \cite{dbmo95})
should be studied and interpreted with great care,
since in all cases we do not know 
the exact locations of substructure clumps
relative to the cluster center along the line of sight,
which could be anywhere from zero to $\sim 10-30h^{-1}$Mpc.
To make it somewhat more quantitative, let us again
take the cluster shown in Figure 1 as an example.
An incorrect naive interpretation would
imply a merger rate $\sim 5$ times too high.

What we have learned here is
that projection effects on X-ray maps are 
complex and easily misleading if careless attempts are made to
interpret observations of X-ray clusters of galaxies.
Furthermore, in order to correctly model such projection effects,
large-scale hydrodynamic simulations with box
size of at least $100h^{-1}$Mpc are required
(both to model the gravitational tidal field
and to capture the local cluster environment properly, noting
that clustering properties in a simulation are only trustworthy
up to the scale about a quarter of the box size).
In the pioneering work by Richstone, Loeb
and Turner (1992), 
and Evrard and co-workers (\cite{emfg93}; \cite{mefg95}),
attempts are made to make connections between
cosmology and substructure/morphology.
It will be very fruitful
to continue this line of investigation
using both improved theoretical modelling with
larger hydrodynamic simulations and improved observations
with higher spectroscopic resolution 
(reducing contamination due to objects at larger distances 
from the clusters).
However, until we can measure real distances of X-ray sources
and galaxies accurately,
we are stuck with the fact that complex motions in the
vicinity of clusters prevent us from fixing the relative
true distances between structures along the line of sight.
This fact dictates that
for some observables
the only meaningful way to compare predictions of 
a cosmological model with the cluster observations 
is to subject clusters in a simulated universe 
to exactly the same observational biases and uncertainties,
and to compare the ``observed" simulated clusters
with real ones.

\subsection{Statistical Analysis --- Quantitative Results} 

Now we turn to a statistical analysis of a sample of 
150 clusters, identified along the three faces of the simulation
box (see \S 2.2).
We divide the clusters into two sets
according to the observed velocity dispersion within the indicated
radius $R_{proj}$; 
we denote the set with $\sigma_{proj}>600$km/s
as ``clusters"
and the set with $300$km/s$<\sigma_{proj}<600$km/s as ``groups",
where $\sigma_{proj}$ is the velocity dispersion 
of a cluster as it would be observed
(after $3\sigma$ clipping).
$R_{proj}$ is the projected cluster-centric radius 
within which anaylses of various quantities are performed.
Three choices of 
$R_{proj}=(0.5,1.0,2.0)h^{-1}$Mpc are used
to examine the dependence of the various projection effects
on the cluster-centric distance.

\subsubsection{Projection Effect on Cluster Richness}

Figure 11 shows the cumulative
probability distribution of the ratio of 
observed number of galaxies to the true number of galaxies (see \S 2.3),
$n_{proj}/n_{clust}$, for ``clusters" [Panel (a)]
and ``groups" [Panel (b)]
at four different projection radii, 
$R_{proj}=(0.5,1.0,2.0)h^{-1}$Mpc
and $R_{proj}=R_{200}$.
$R_{200}$ is the radius within which the mean density
of the cluster is 200 times the critical density,
and is individually computed for each cluster.
$R_{200}$ approximately indicates
the boundary of the virialized region of a cluster
(Gunn \& Gott 1972).

The median values of the distributions
for the ``clusters" and ``groups"
are comparable with
$(n_{proj}/n_{clust})_{median}\approx (1.10,1.20,1.60)$ 
at $R_{proj}=(0.5,1.0,2.0)h^{-1}$Mpc, respectively,
i.e., on average, the richness of
each cluster is overestimated by about ($10\%, 20\%, 60\%$)
within the three chosen radii.
Only $(2\%, 2\%, 20\%)$ of the clusters have
$n_{proj}/n_{clust}>2.0$ 
at $R_{proj}=(0.5,1.0,2.0)h^{-1}$Mpc.
If one corrects for mean background density of galaxies,
the small overestimate in richness will be further reduced.
Note that the results with $R_{proj}=R_{200}$ 
and $R_{proj}=1.0h^{-1}$Mpc for the clusters
are very similar, indicating that the virial radius
for the clusters considered is about $1.0h^{-1}$Mpc.
The same argument implies that the virial radius for
the groups is about $0.75h^{-1}$Mpc.
It appears that the richness of a cluster is not
significantly altered by projection at a radius
$R_{proj}\sim 1.0h^{-1}$Mpc.

We may interpolate between the dotted and short dashed curves
in Panel (a) of Figure 11 to infer the possible richness contamination
within Abell radius $R_{Abell}=1.5h^{-1}$Mpc.
Approximately 30\% of the clusters have 
their richness overestimated by about 50\% or more at $R_{Abell}$.
Note that our simulated ``clusters" are selected out using a much more
strict criterion than that used to select Abell clusters.
In other words, real Abell clusters would suffer from
significantly more contaminations.
In this sense, our results are consistent with those of
Frenk \etal (1990; their Table 2),
who find that about 50\% of Abell clusters 
are not real three-dimensional rich clusters
in their bias $b=2$ CDM model which best
reproduces the large-scale and clusters observations among
their models; rather, they 
are projected configurations of groups or poor clusters 
aligned in the radial directions.

Usually, ``groups" have higher probabilities
of being contaminated than
``clusters" at large $n_{proj}/n_{clust}$.
For example, the richness of only $5\%$ of clusters is
overestimated by $50\%$ or more, whereas that of 13\% of
groups is overestimated by $50\%$ or more 
at $R_{proj}=1h^{-1}$Mpc.
Part of the reason is that clusters are intrinsically richer,
so the relative contamination is smaller,
although the absolute amount of contamination is larger on average
for clusters than for groups.
We may extrapolate that the contamination will be
smaller for still richer clusters.
This trend is also found in Frenk \etal (1990; Table 2) for
Abell-like clusters in that a low bias
(a high amplitude) model tends to have a smaller fraction
of clusters due to projection than a high bias model.

\subsubsection{Projection Effect on Cluster X-ray Luminosity} 

Figure 12 shows the cumulative
probability distribution of the ratio of 
observed cluster X-ray luminosity to the true cluster X-ray
luminosity due to the hot gas {\it in the cluster},
$L_{x,proj}/L_{x,clust}$, 
in the 0.4-2.4keV band (a) and in the 0.5-10.0keV band (b).
Four cases are shown for ``clusters" and for ``groups" 
at four different projection radii 
$R_{proj}=(0.5,1.0,2.0)h^{-1}$Mpc
and $R_{proj}=R_{200}$.

We see an overestimate of the true X-ray luminosities 
by (2\%, 8\%, 20\%) 
at $R_{proj}=(0.5,1.0,2.0)h^{-1}$Mpc, respectively,
(the effects for the ``groups" and ``clusters" are
comparable at the same radius) 
in the 0.4-2.4keV band,
which should be compared with a (10\%, 20\%, 60\%) effect
for galaxy number contamination (Figure 11).
The fact that X-ray emission is a much stronger function of 
density and temperature 
(bremsstrahlung emission 
$e_{x}\sim\rho^2 T^{1/2}\exp^{-hv/kT}$) than the 
optical counterpart, as normally considered to be one of the
advantages of X-ray observations of clusters of galaxies
against background/foreground contaminations,
is indeed borne out from this analysis.
We conclude
that the projection effects on X-ray luminosity
of a cluster is small at radii $\leq 1.0h^{-1}$Mpc,
typical for X-ray observations.
This finding makes the cluster X-ray luminosity function 
a very useful and simple diagnostic for comparing observations
(Henry \& Arnaud 1991; Henry 1992)
with theoretical predictions
(Kang \etal 1994; Bryan \etal 1994; 
Cen \& Ostriker 1994). 

We see that the projection effect is comparable 
and perhaps only slightly smaller 
in the harder 0.5-10.0keV  X-ray band than in the 0.5-2.4keV band.
This result should be interpreted more carefully.
Since the bulk of our clusters have richness zero or one,
and thus relatively low temperatures,
the  two adopted X-ray bands therefore do not differ
significantly.
We expect that for richer, hotter clusters (kT$\ge 5$keV)
the advantage of a harder and wider band should be much more visible.

In general, the projection effect on X-ray luminosity of a cluster
increases with radius, as expected.
We speculate that heavily projection contaminated clusters
may show significantly shallower baryonic density profiles than
their total mass counterparts,
due to the fact that projection effects
are more significant in the outer part than in the 
inner part of a cluster and that, while projection on average
causes underestimation of the cluster mass, it can only
causes overestimation of the baryonic mass.
There may be some observational evidence for this speculation
(Markevitch \etal 1996), which is further strengthened
by another finding that gas density
profile and total density profile 
are nearly parallel to one another 
when no projection is allowed (Frenk \etal 1996).

\subsubsection{Projection Effect on Velocity Dispersion and Mass Estimates} 

We now examine how projection effects alter velocity dispersion and
various mass estimates for a cluster.
We calculate 
the true cluster mass (computed by counting
all the particles within the indicated 3-d radius)
and three observational mass estimates: 
the isothermal virial mass estimate, the X-ray isothermal hydrostatic 
equilibrium mass estimate,
and the gravitational lensing mass estimate.

Figure 13 shows the cumulative
probability distribution of the ratio of the 
observed 1-d velocity dispersion to the true 1-d velocity 
dispersion calculated by
considering the true cluster members only, 
$\sigma_{proj}/\sigma_{clust}$.
Four cases are shown
for ``clusters" and for ``groups" 
at four different projection radii 
$R_{proj}=(0.5,1.0,2.0)h^{-1}$Mpc
and $R_{proj}=R_{200}$.
We see that, on average, projection
causes an overestimate of the true velocity dispersion by
(5\%, 9\%, 27\%) for clusters,
and (2\%, 4\%, 10\%) for groups
at $R_{proj}=(0.5,1.0,2.0)h^{-1}$Mpc, respectively.
It appears that the velocity dispersions of clusters
are only slightly overestimated due to projection.
About 10\% of clusters have their
velocity dispersions overestimated by about 40\%.
This contrasts with the Abell clusters, 
for which Frenk \etal (1990; their Figure 3b)
find that 10\% of Abell clusters
with the highest velocity dispersions found in the simulations
are severely contaminated due to projection, causing an overestimate of
true cluster velocity dispersions by at least 70\%.

Figure 14 shows the cumulative
probability distribution of the ratio of derived
cluster mass assuming isothermal velocity dispersion [using 
the average 
velocity dispersion within the indicated $R_{proj}$:
$M_{VT}\equiv 2\sigma_{proj}^2(<R_{proj}) R_{proj}/G$]
to the true mass within the indicated radius, 
$M_{VT}/M_{clust}$.
Four cases are shown 
for ``clusters" and for ``groups" 
at four different projection radii 
$R_{proj}=(0.5,1.0,2.0)h^{-1}$Mpc
and $R_{proj}=R_{200}$.
Figure 15 shows 
$M_{VT}/M_{clust}$ as a function of $M_{clust}$ 
for clusters (solid dots) and groups (open circles).

An examination of Figures (14,15)
reveals several interesting points.
First, 
the projection effect on mass estimates differs
strongly between $R_{proj}=1$ and 2$h^{-1}$Mpc.
The median value of $M_{VT}/M_{clust}$
is (1.10,1.11,1.59) for ``clusters" and
(0.75,0.78,1.02) for ``groups"
at $R_{proj}=(0.5,1.0,2.0)h^{-1}$Mpc, respectively.
Second, in general, the isothermal mass estimates of groups 
at $R_{proj}\le 1h^{-1}$Mpc (typically used in observations)
underestimate the true masses by about 25\%, on average, 
with the $2\sigma$ lower and upper limits being (0.50,1.20)
(noting the non-normal distribution),
and there is no correlation 
between  $M_{VT}/M_{clust}$ and $M_{clust}$ for groups.
Third, cluster masses are, on average,
overestimated by about 
10\% at $R_{proj}\le 1.0h^{-1}$Mpc using the isothermal model,
with the $2\sigma$ lower and upper limits being (0.60,1.90)
(a much broader distribution than that of the groups).
A large radius such as $R_{proj}=2.0h^{-1}$Mpc
causes fairly large overestimates (with a very broad distribution) 
of the true cluster masses, and therefore is not a suitable choice
(this observation is consistent with the fact
that the virial radius $\sim 1.0h^{-1}$Mpc
is much smaller than $2.0h^{-1}$Mpc
for the clusters under examination).
However, we note that the tail of high 
$M_{VT}/M_{clust}$ for ``clusters" is caused by
clusters which
have intrinsically lower masses but are severely contaminated in
velocity space.
These clusters are categorized as ``clusters" precisely 
because of their inflated, observed velocity dispersions.
Finally, we point out that,
for the most massive clusters 
[$M(<1.0h^{-1}Mpc)\ge 3\times 10^{14}h^{-1}\msun$], 
the situation is different from that for the entire cluster set.
At $R_{proj}=1.0h^{-1}$Mpc, the most massive clusters'
masses are, on average, {\it underestimated} by about 20\%
by the isothermal model.
Perhaps by coincidence, the isothermal model
gives, on average,
correct estimates for the masses of the most massive clusters
at $R_{proj}=2.0h^{-1}$Mpc.
An extrapolation based on 
the $\sim 10$ rightmost points in both
Panels (b) and (c) of Figure 15
indicates that 
clusters more massive than those contained
in the simulation box would perhaps suffer from mild underestimates 
of their true masses by
the isothermal model by about 10-30\% at $R_{proj}=1.0-1.5h^{-1}$Mpc.
This extrapolation deserves further study with larger simulations,
especially since richer, more massive clusters are 
more accessible observationally.

Figure 16 shows 
$M_{VT}/M_{clust}$ as a function of $n_{proj}/n_{clust}$ 
for clusters (solid dots) and groups (open circles).
Both clusters (solid dots) and groups (open circles) show no correlation
between the two plotted quantities,
but there is a clear trend of larger
scatter towards larger values in the x-axis.

We now switch gears to investigate
the X-ray determination of cluster masses.
Figure 17 shows the cumulative
probability distribution of $M_{xray}/M_{clust}$, 
where $M_{xray}$ is the mass derived 
from X-ray observation of the luminosity-weighted
temperature  of a cluster
[within the indicated radius:
$M_{xray}=2k<T>_{emis} R_{proj}/(\mu m_p G)$,
where $\mu=0.60$ is the molecular weight,
$<T>_{emis}$ is the emissivity-weighted
temperature within the projected radius,
and other symbols have their usual meaning],
assuming that the cluster is
in isothermal hydrostatic equilibrium.
Figure 18 shows
$M_{xray}/M_{clust}$ as a function of $M_{clust}$.
We see that X-ray mass estimates are relatively larger
for groups than for clusters, a trend
opposite to what is found in Figure 14 for
galaxy velocity dispersion derived mass estimates,
due to the fact that some of the 
intrinsically poor but velocity-inflated (due to projection)
clusters do not have inflated temperatures.
This difference is worth noting:
the projected luminosity-weighted
X-ray temperature is, to some extent, independent upon
the relative motions among the X-ray sources, barring ongoing mergers.
At $R_{proj}=0.5h^{-1}$Mpc,
X-ray derived masses underestimate, on average,
the true masses by about 12\% for both clusters and groups.
At $R_{proj}=1.0h^{-1}$Mpc,
X-ray derived masses underestimate
the true masses by about 15\% and 8\% 
for clusters and groups, respectively.
But they overestimate
the true masses by about 3\% and 14\% for clusters and groups, respectively,
at $R_{proj}=2h^{-1}$Mpc.
Comparing with Figures 14,15 it appears that
X-ray mass determinations are relatively more stable 
and have smaller dispersions. 
Noticeably, several solid dots on the upper left corner
in the Panels (a,b,c) of Figure 15 
have moved down in Figure 18.

Finally, we add that for the most massive clusters  in the simulation box
[$M(<1.0h^{-1})$Mpc$\ge 3\times 10^{14}h^{-1}\msun$]
the X-ray derived masses are comparable 
to those derived using the galaxy velocity dispersions,
with both fairly robustly
underestimating the true masses by about 10-30\%;
this finding is echoed by the fact that real observed
clusters also show such an agreement 
between the two kinds of masses (Bahcall \& Cen 1993).

Evrard, Metzler, \& Navarro (1996; EMN hereafter) 
conclude that, on average,
the X-ray temperature determined mass agrees 
remarkably well with the true cluster mass.
We attribute a large part of the difference 
between our results here and theirs 
($M_{xray}/M_{clust}\sim 0.80$ in 
this study versus $M_{xray}/M_{clust}\sim 1.0$ in EMN) 
to the difference in simulation box size.
We note that 52 out of 58 clusters are simulated using boxes with sizes
$\le 30h^{-1}$Mpc in EMN.
As we find previously by visually examining 
the clusters that projection effects on a
cluster along the line of sight are mostly contributed by objects with 
line of sight distances from the cluster $\pm 30h^{-}$Mpc. 
Let us assume conservatively for the purpose of illustration
that $\pm 15h^{-1}$Mpc bracket the relevant region.
Since a simulation with periodic boundary conditions 
is only trustworthy on scales
up to about a quarter of the box size, this 
implies that a box size of at least 
$4\times (15+15) = 120 h^{-1}$Mpc 
is needed in order to properly allow for 
possible projection effects. 
Note that the box size of the 
simulation analyzed here has a size of $128h^{-1}$Mpc. 
The remaining 6 clusters analyzed by EMN are from
Navarro, Frenk \& White (1994; NFW hereafter), which properly include
the large scale tidal field using a box size of $180h^{-1}$Mpc
as well as hydrodynamics.
In fact, examining Figures (6,7) of EMN reveals
that the 6 NFW clusters on average show
$M_{xray}/M_{clust}\sim 0.8-0.9$ on the scale where
the mean density is about $200\rho_c$ (corresponding approximately
to our $R_{proj}=1h^{-1}Mpc$),
in good agreement with what is found here.

On the other hand,
in a recent work Bartelmann \& Steinmetz (1996)
find that X-ray determined cluster mass by fitting
emission profiles using $\beta$ model
underestimates the true mass by about 40\%
within the radius where overdensity is 500 relative to the 
global mean.
Since our simulation prevents us from obtaining 
reliable results at an overdensity of 500,
it is not possible to make a one to one comparison,
but it seems that
their results are not inconsistent with results
obtained here since there are three major differences between
the two studies:
1) they use a detailed gasdynamic simulation although the resolutions
in the two studies are comparable,
2) emission profiles may not have reached their asymptotic
slopes at an overdensity of 500, as they have indicated,
3) the treatments of projection effects are different.

Next, we examine
the gravitational lensing mass estimate.
Figure 19 shows 
the cumulative probability of $M_{lensing}/M_{clust}$. 
$M_{lensing}$ is the total mass including 
all the matter within a line-of-sight
distance of $64h^{-1}$Mpc from the cluster within
the indicated projection radius $R_{proj}$,
subtracted off by the mean background mass inside such a volume.
In other words, we assume conservatively
that techniques such as weak lensing
mass reconstruction can do a perfect job to recover the mass
inside the beam.
We see that gravitational lensing mass is, on average,
overestimates the true mass by only 5-10\%.
However, the dispersion is comparable to that
of the velocity dispersion determined mass but
much larger than that of the X-ray temperature determined mass.
Figure 20 shows 
the cumulative probability of $M_{lensing}/M_{VTc}$. 
inside the beam.
$M_{VTc}$ is the virial (isothermal) mass estimate 
using the cluster velocity dispersion,
corrected to its 2-d projected
value (using the 3-d density profile of each
individual cluster outside the indicated radius $R_{proj}$; i.e.,
$M_{VTc}$ includes mass outside $R_{proj}$ along the line-of-sight).
$M_{lensing}$ is the mass that 
an analysis of lensing observations would derive,
assuming that the additive quantity due
the ambiguity in the mass surface density near the
edge of an observed galaxy sample in the methods
of weak lensing mass reconstruction 
(\cite{tvw90}; \cite{m91}; \cite{ks93}) can be properly calibrated.
In real observations, the lensing mass estimate 
may be subject to more projection than used here,
since any mass along the line of sight may contribute
to the surface mass density (although with varying weights
depending on redshift).
Figure 21 shows $M_{lensing}/M_{VTc}$ as a function 
of $M_{clust}$.
The first noticeable feature in the gravitational
lensing mass to virial mass ratio
is that groups are severely contaminated, causing a large 
scatter in $M_{lensing}/M_{VTc}$ (Figure 21)
and consequently a much broader distrbution
of $M_{lensing}/M_{VTc}$ in Figure 20.
Secondly, $M_{lensing}/M_{VTc}$ for the most massive clusters 
[$M(<1.0h^{-1}Mpc)\ge 3\times 10^{14}h^{-1}\msun$]
is $\sim 1.3$ at $R_{proj}=1.0h^{-1}$Mpc and
$\sim 1.2$ at $R_{proj}=2.0h^{-1}$Mpc.

It is also interesting to compare the lensing derived mass
to X-ray temperature derived mass.
Figure 22 shows 
the cumulative probability of $M_{lensing}/M_{xrayc}$, 
where $M_{xrayc}$ is the X-ray derived mass
within the 2-d projected radius $R_{proj}$.
Figure 23 shows $M_{lensing}/M_{xrayc}$ as a function of $M_{clust}$.
Figures 22,23 are similar to Figures 20,21,
although $M_{lensing}/M_{xrayc}$ has a slightly tighter 
distribution than that of 
$M_{lensing}/M_{VTc}$ for the most massive clusters
[$M(<1.0h^{-1}Mpc)\ge 3\times 10^{14}h^{-1}\msun$].
At $R_{proj}=1.0h^{-1}$Mpc, the
derived values of $M_{lensing}/M_{xrayc}$
for the most massive clusters 
are fairly consistent with those for $M_{lensing}/M_{VTc}$;
the median value of $\sim 1.3$ has
a surprisingly small dispersion of $\sim 0.2$.

It is helpful to compare the various mass estimates obtained above
in a more systematic fashion to understand the
inter-relationship among them.
The read is invited to examine Figures (15,18,21,23)
more closely.
We conclude that no clear correlation between
the quantity shown in the ordinate and the true
cluster mass ($M_{clust}$) is visible in all four figures.
It implies that the various mass ratios are also uncorrelated.
This property is common 
in complex multivariate systems (Kendall 1980),
such as clusters of galaxies examined here. 
Some consequences should be noted.
For example, histograms of the various mass ratios
may appear contradictory or are not fully translatable;
i.e., the widths and median or average values of the distributions
are easily related.
In other words, taking ratios is not a
linear operation so the propagation of mean (or median)
values does not necessarily behave in a simple way.

A solid statistical comparison of the results
shown in Figures (20,21,22,23) with observations 
would require more careful analysis of the theoretical
models in two major ways as well as larger 
homogeneous observational samples.
First, large-scale hydrodynamic simulations
which properly compute gasdynamic effects are needed.
What we have done here is approximate, especially for
the temperature of a cluster, which in general does not 
need to be equal to the dark matter temperature (square of its
velocity dispersion).
Second, one should employ the exact same methods
used observationally to derive lensing, virial and X-ray masses
to analyse simulations,
utilizing real density,
velocity and temperature distributions 
rather than assuming isothermal (hydrostatic) distributions.
Nevertheless, some crude comparisons
between the simulations and existing observations
may be informative.

Observations of the Abell cluster A1689 seem to show
that the lensing derived mass is much lower 
than that
using the galaxy velocity dispersion data 
($\sigma_{proj}=2355km/s$)
within a radius
of $\sim 1.0h^{-1}$Mpc (Tyson \& Fischer 1995),
while our simulation indicates
that the lensing derived mass is usually 
larger than the dynamically derived mass.
So the situation is intriguing.
It implies that perhaps the velocity dispersion
of A1689 is severely contaminated and inflated.
There is some evidence that this may indeed 
be the case for A1689 (Teague \etal 1990).
The hint is that 
 there are several 
components about 10,000-15,000km/s away from the cluster center 
(last panel of Figure 5 of Teague \etal 1990),
which are, of course, not included in the velocity dispersion estimate.
It seems plausible that some other more nearby
structures may have merged
into the main cluster's velocity structure and
happen to constitute
the tails of the total observed cluster velocity distribution.
Probably a more reliable
measure of the potential well 
is provided by the X-ray observation (Yamashita 1994),
$k T_{x}=(7.6\pm 0.5)$keV, which, in the case of $\beta=1$,
translates into a 1-d velocity dispersion of 854km/s. 
This is 2.8 times smaller than
the observed galaxy velocity dispersion,
and is in better agreement with lensing observations.

Another example is Abell cluster A2218 at redshift z=0.175,
which has a gravitational weak lensing derived mass of 
$(3.9\pm 0.7)\times 10^{14}h^{-1}\msun$ within
$\sim 400h^{-1}$kpc radius (\cite{s96})
and an X-ray temperature derived mass 
(assuming the intracluster gas being in hydrostatic equilibrium) 
$(2.6\pm 1.6)\times 10^{14}h^{-1}\msun$.
The ratio $M_{lensing}/M_{xrayc}\sim 1.5\pm 1.2$.
Within the observational uncertainties the two mass estimates formally
agree with one another.
On the other hand, 
$M_{lensing}/M_{xrayc}\sim 1.5$ can be easily 
reconciled by the projection effects.

Analysis of the relatively poor cluster MS 1224+20 at redshift z=0.325
by Carlberg, Yee, \& Ellingson (1994)
results in an 
$M_{lensing}/M_{VTc}$ ratio of $2.5\pm 1.1$. 
This same cluster is also analysed by 
Fahlman \etal (1994) using the Kaiser-Squires (1993) algorithm,
yielding a value of 3.0 for 
$M_{lensing}/M_{VTc}$, consistent with the result
of Carlberg \etal (1994).
From Figure 21 we note that it is not uncommon for poor 
clusters to exhibit large $M_{lensing}/M_{VTc}$ ratio;
a value 2.5 is well within the scatter.
It should be noted that clusters 
out of dynamical equilibrium 
do not necessarily give larger virial mass estimates.

Another Abell galaxy cluster,  A2390 at redshift z=0.23,
is analysed by Squires \etal (1996).
They show that
the lensing derived mass is higher than that derived
from velocity dispersion by a factor of $\sim 1.6$ at a radius
of $R_{proj}\sim 0.7h^{-1}$Mpc (taking
the rightmost point
in Figure 3 of Squires \etal 1996 and 
comparing it to the dashed curve at $\theta\sim 260$ arcseconds).

Smail \etal (1995) analysed two clusters
Cl 1455+22 at redshift z=0.26 and
Cl 0016+16 at redshift z=0.55,
finding that the lensing deduced masses are in good
agreement with those derived using other methods.

To summarize, the observed clusters, which have
both lensing derived mass and either virial mass or 
X-ray mass, in general, show 
$M_{lensing}/M_{VTc, xrayc} > 1.0$, except for A1689.
It seems that projection effects on the various quantities
involved may well account for these numbers at present time
given the large statistical uncertainties.
Furthermore, a modest (20\%) amount of velocity bias of galaxies over
dark matter 
(Carlberg, Couchman \& Thomas 1990; Carlberg \& Dubinski 1991; 
Cen \& Ostriker 1992; Evrard, Summers, \& Davis 1994;
Brainerd \& Villumsen 1994; 
see also Katz, Hernquist \& Weinberg 1992)
would raise the 
computed ratio of the lensing to dispersion-based mass in 
the simulation by a factor of 1.4. 
Similar temperature bias has the same effect. 
However, if the relatively
large values of
$M_{lensing}/M_{VTc, xrayc}\ge 3.0$ 
survive with future improved observations with lower uncertainties, 
we will be perhaps forced to
re-examine our understanding of the physical and 
dynamical processes in the clusters of galaxies
(e.g., Loeb \& Mao 1994; Miralda-Escud\'e \& Babul 1995).

Finally, we would like to note that it is anticipated that
the projection effects will become progressively more severe
at higher redshift because 1) 
structures in the past tend to be more filamentary and sheet-like
(i.e., less knoty),
and 2) observations become more difficult
at higher redshift.

\subsubsection{Gas-to-Total Mass Ratio in Clusters}

We now study the issue of gas to total mass ratio
in clusters of galaxies, a quantity of major cosmological 
importance, recently emphasized by White \etal (1993).
It is important 
because it places an important constraint on $\Omega_b/\Omega_{tot}$,
where $\Omega_b$ and $\Omega_{tot}$ is the density of baryons
and total density of non-relativistic matter in units of
critical density.
Figure 24 shows 
the differential probability distribution of the
gas to total mass ratio $M_{gas}/M_{tot}$,
for ``clusters" (thick curve) and ``groups" (thin curve)
at a radius $R_{proj}=1.0h^{-1}$Mpc,
in units of $\Omega_b/\Omega_{tot}$.
We take $M_{tot}=M_{xray}$ in this case.
In taking account the projection effects 
we simply say that the true intrinsic ratio 
(which is assumed unity in units of the global ratio of
the baryonic density to total density)
will be modified by the change in the derived gas mass
and in the total mass estimate as:
$M_{gas}/M_{tot}=(M_{gas,proj}/M_{gas,clust})/(M_{xray}/M_{clust})$
(see Figures 18,19 for the distribution of $M_{xray}/M_{clust}$),
where $M_{gas,proj}$ is the total hot ($T>10^6$Kelvin)
baryonic mass projected within $R_{proj}=1.0h^{-1}$Mpc
and 
$M_{gas,clust}$ is the instrinsic hot ($T>10^6$Kelvin)
cluster baryonic mass  projected within $R_{proj}=1.0h^{-1}$Mpc
(the gas mass is obtained by simply multiplying
the corresponding total mass by $\Omega_b/\Omega_{tot}$).
In the calculation of  
$M_{gas,proj}$ and $M_{gas,clust}$
only particles with $\rho \ge~<\rho>$ and temperature greater
than $0.32$keV are included.
A more sophisticated treatment by de-projecting the surface
brightness distribution to obtain the gas mass 
(e.g., Fabricant, Rybicki, \& Gorenstein 1984)
is not attempted here due to the rather approximate
treatment of the X-ray maps.
Figure 25 shows $M_{gas}/M_{tot}$ as a function  of $M_{clust}$.

On average, clusters and groups
have a value of 
$M_{gas}/M_{tot}$ larger than the
global ratio by 30-40\%.
This helps but is not quite sufficient to 
reconcile the standard nucleosynthesis value
of $\Omega_b=0.0125h^{-2}$ (Walker \etal 1991)
and $\Omega=1$ 
with the observed gas to mass ratio value in
clusters of galaxies, $0.05h^{-3/2}$ (Lubin \etal 1996), 
for any reasonable $h$. 
For example, if we take a 40\% increase due to projection effects,
an $h=0.7$, $\Omega=1$ model would have on average
the gas to total mass ratio of $0.036$
in clusters, still a factor of 2.4 smaller 
than $0.085$, the value observed in real clusters for the given $h$.
However, the broadness of the observed distribution 
(Figure 4 of Forman \& Jones 1994; Lubin \etal 1996) of 
$M_{gas}/M_{tot}$ or large variations in baryon fraction
from cluster to cluster (or group to group) (Loewenstein \& Mushotzky 1996)
seems to be adequately accounted for by the
projection effects without the need to invoke 
other processes such as
gas being driven out of the clusters 
due to supervovae explosion energy
deposited in the cluster gas, 
or gas being more segregated
in clusters than dark matter by some yet unknown processes.
Furthermore, groups tend to occupy more of the low end
of the distribution, again as indicated by observations 
(Forman \& Jones 1994).
Finally, the null correlation between 
$M_{gas}/M_{tot}$ and $M_{clust}$ (or velocity dispersion),
shown in Figure 25 and consistent with observations,
hints that any segregation process between
gas and dark matter which strongly depends on the mass or potential
depth of a cluster, for the clusters 
in mass range that we have examined,
does not play a dominant role.

The origin of the moderate projection effect
on the gas to total mass ratio is the following.
Roughly speaking,
the overall effect consists of two separate effects.
The first effect is that
the cluster masses are underestimated by about 10-20\%
(see Figures 18,19) for the cluster case.
The second effect is 
that the cluster baryon mass is overestimated by about 10-20\%
for the cluster case (not shown),
due to projected background 
and foreground hot gas. 
The latter effect is somewhat similar 
to the lensing mass overestimate shown in Figure (20) but
differs in 
that the lensing mass has been subtracted off by the background mean
mass, while for the hot gas mass no such subtraction is performed.

Finally, we note that
real three-dimensional (adiabatic or with {\it positive} feedback)
hydrodynamical
simulations have consistently shown that
the baryons in the clusters 
are actually {\it anti-biased} relative to the matter. 
It has been found that the baryon to matter ratio
in three-dimensional clusters is approximately 0.90 in units of the
global ratio 
(Figure 14 of Cen \& Ostriker 1993a; White \etal 1993;
Evrard \etal 1994).
This effect, unaccounted for in our treatment,
would slightly shift leftward the curves in Figure 24,
somewhat compensating the projection effects.
But the situation is somewhat more delicate.
The antibias of gas to mass of $0.90$ found in three
dimensional clusters is presumably because
some fraction of gas is driven out of the cluster by shocks.
Therefore, this shock-driven-out gas is hot and some fraction of it along
the line of sight would show up in cluster X-ray maps.
So a more realistic multiplicative factor might be $0.9-1.0$,
depending on the distance of the shock front from the cluster center,
the thermal and dynamic history and the geometry of a cluster,
observational beam size and sensitivity,
and foreground/background contamination.
Furthermore, non adiabatic treatment of the intracluster
gas, i.e., including cooling, 
might reduce the amount of anti-bias found in the 
adiabatic simulations.
On the other hand, feedback processes associated
with galaxy formation might have opposite effects.
All these considerations indicate that
large hydrodynamic simulations are invaluable to 
settling this important issue.

\subsubsection{Projection Effect on Substructure} 

Lastly, the projection effects on substructure
in clusters of galaxies are examined.
Here no attempt is made to examine all proposed substructure 
measures (for a complete list of substructure measures
see Pinkney \etal 1996).
We pick two measures for analysis of galaxy maps and one for X-ray maps.
We begin with an analysis of the velocity distribution,
parameterized by the kurtosis, ``$Kurt$".

Figure 26 shows the distribution of the difference between
two kurtoses: $Kurt_{clust}$ is the intrinsic kurtosis
of the cluster velocity distribution without any 
projection contamination, 
$Kurt_{proj}$ is the kurtosis of the total
cluster velocity field including
projected foreground and background structures.
Figure 27 plots $Kurt_{proj}$ against $Kurt_{clust}$.
Three points are interesting.
First, the intrinsic kurtosis distribution itself
is rather broad, i.e., many groups or clusters do
not possess Gaussian velocity 
distributions ($Kurt=0$) even without projection contamination.
Second, projection effects change the intrinsic kurtosis
of a cluster or a group rather randomly at large projection
radii of $R_{proj}=2.0h^{-1}$Mpc;
at smaller projection radii, there is a correlation between
$Kurt_{proj}$ and $Kurt_{clust}$, albeit with large dispersions.
Finally, given the rather large scatter seen in Figure 27,
it seems that kurtosis of the velocity field of a galaxy cluster
perhaps does not serve as
a good measure of substructure in a cluster.

Another statistic, designed to measure substructures
which are localized in both spatial and velocity spaces,
is the Dressler-Shectman $\Delta$ statistic (Dressler \& Shectman 1988):
\begin{equation}
\Delta = \sum_{i=1}^N \delta_i,
\end{equation}
where $N$ is the total number of galaxies in the cluster,
and $\delta_i$ for
each galaxy is defined as:
\begin{equation}
\delta_i^2 = (11/\sigma^2)[(\bar v_{local}-\bar v)^2 + (\sigma_{local}-\sigma)^2],
\end{equation}
where $\bar v$ and
$\sigma$ 
are the mean velocity and global velocity dispersion of the cluster;
$\bar v_{local}$ and
$\sigma_{local}$ 
are the local mean velocity 
and local velocity dispersion 
for 10 nearest neighbors around a galaxy. 
Following Dressler \& Shectman (1988), 1,000 Monte Carlo 
models are run to calibrate the $\Delta$ statistic for
each cluster. Each Monte Carlo model is made by randomly
shuffling the velocities among the cluster galaxies.
Then, we define $P(>\Delta)$ as the fraction of the total
number of Monte Carlo models of the cluster that have
$\Delta's$ larger than the true value of the cluster.
$P(>\Delta) \sim 1.0$ means that the cluster contains no
substructure, while 
$P(>\Delta) \sim 0.0$ indicates that the cluster contains 
statistically significant substructure.

Figure 28a shows the probability distribution of 
$P(>\Delta)$ for the ``true" cluster without projected external structures,
at three projection radii $R_{proj}=(0.5,1.0,2.0)h^{-1}$Mpc.
The probability is normalized such that the sum of all
the bins is unity.
Figure 28b shows the corresponding probability distribution of 
$P(>\Delta)$ for the actual observed cluster including
projected galaxies.
The projection effect is {\it striking}.
We see that only 
(4\%,0\%), (4\%,2\%), (7\%, 1\%)
of (clusters,groups) have
intrinsically, statistically significant ($P_{clust}<0.1$) 
substructures at $R_{proj}=(0.5,1.0,2.0)h^{-1}$Mpc, respectively,
when only the true members are analyzed 
(the uncertainty is about $\pm 3\%$ due to random sampling
of the particles).
In sharp contrast,
(53\%,45\%), (72\%,63\%) and (95\%,85\%) of (clusters,groups)
at $R_{proj}=(0.5,1.0,2.0)h^{-1}$Mpc, respectively,
are found to show statistically significant substructures,
when the clusters are actually ``observed"
(the uncertainty here is about $\pm 10\%$ due to random sampling
of the particles).
The large fraction of ``observed" clusters of galaxies 
having significant substructure
in this cosmological model 
is in accord with
that derived from real observed clusters of galaxies
(Dressler \& Shectman 1988).
Geller \& Beers (1982) use a different technique to 
study the substructure in clusters and reached the same conclusion
about the fraction of observed clusters with substructure.
The trend of more substructure with larger projection
radii is consistent with that found by
West, Oemler, \& Dekel (1988) and West \& Bothun (1990).
Having found agreement between observed clusters
and simulated clusters analyzed in the same way,
the fact that true clusters contain much less substructure
implies that 
{\it the large amount of 
substructure in most of real observed clusters of galaxies
is perhaps due to projection effects}.

We now turn to the X-ray maps to examine the 
projection effects on substructure
[see Forman \& Jones (1994) for an excellent
review of the current status of observations on
the subject of substructure in X-ray clusters].
Here a simple measure is used to measure substructure 
in X-ray maps, which was invented by Davis \& Mushotzky (1993) 
in analyzing Einstein imaging data. 
The measure is defined as follows.
We find all the local maxima in an X-ray surface
brightness map, and define the luminosity of each 
local maximum as the sum of luminosities of all the pixels
``associated" with the local maximum.
``Associated" pixels are found by propagating each pixel
along the gradient of the surface brightness 
until it reaches a local maximum (where the gradient is zero).

Figure 29a shows the distribution of 
$N(L_{sub}>0.1L_{main})$, the number of substructures
each having a luminosity larger than 10\% of that
of the main (central) cluster structure in the 0.4-2.4keV band
within three projection radii
$R_{proj}=(0.5,1.0,2.0)h^{-1}$Mpc.
X-ray maps in the 0.4-2.4keV band are used here.
Two curves are shown at each projection radius:
the thin curve is generated from the true cluster X-ray maps
due to intrinsic cluster hot gas only,
the thick curve is for 
the corresponding maps including projected structures.
Figure 29b shows the equivalent distribution 
for groups.
Figures 30a,b are similar to Figures 29a,b,
but only requiring 
that the substructures 
each have a luminosity larger than 1\% (instead of 10\%) of that
of the main structure.
An inspection of 
Figures 29,30 reveals
that the projection effect on the
substructure of X-ray maps is strongly cluster-centric distance
dependent.
The effect at small radius $R_{proj}=0.5h^{-1}$Mpc appears
insignificant:
the fraction of clusters containing no substructure
with luminosities larger than (10\%, 1\%) of that
of the main structure decreases 
(from 84\% to 81\%, from 70\% to 70\%),
respectively, due to projection;
for the groups the numbers change 
(from 86\% to 84\%, from 80\% to 73\%).
But the effect becomes larger at larger radii, especially
when a weak criterion is used:
the fraction of clusters containing no substructure
with luminosities larger than (10\%, 1\%) of that
of the main structure decreases (from 82\% to 78\%, from 54\% to 32\%),
respectively, due to projection at $R_{proj}=1.0h^{-1}$Mpc;
the changes are (from 77\% to 62\%, from 46\% to 11\%),
respectively, due to projection at $R_{proj}=2.0h^{-1}$Mpc;
comparable effects are found for groups.
Furthermore, we note that 
all the cases
of having more than two substructures at $R_{proj}\le 2h^{-1}$Mpc
with $L_{sub}>0.1L_{main}$ are due to projection;
so do all the cases
of having more than three substructures at $R_{proj}\le 2h^{-1}$Mpc
with $L_{sub}>0.01L_{main}$.

Inter-comparison between Figure 28 and Figures 29,30 
indicates that different measures of substructures in galaxy maps
and X-ray maps may yield quite different results.
In general, only a small fraction, 10-20\% of clusters
with radius $R_{proj}=0.5-2.0h^{-1}$Mpc
shows intrinsic substructure, as indicated by
both the Dressler-Shectman $\Delta$ statistic and 
the X-ray surface brightness local maximum measure with $L_{sub}>0.1L_{main}$.
The observed large fraction of clusters showing substructures
in galaxy maps is caused by projection. 
X-ray maps are somewhat more immune to projection 
at small radii $R_{proj}<1h^{-1}$Mpc,
if one demands a sufficiently large $L_{sub}$  
($L_{sub}\ge 0.1L_{main}$, for example).
However, large $R_{proj}\ge 2.0h^{-1}$Mpc
and/or lower $L_{sub}$
($L_{sub}\le 0.01L_{main}$, for example)
are unlikely to be very useful,
if one's goal is to understand the intrinsic substructuring
properties of a cluster.
We note that, although for the clusters considered in this
work [$M(<1h^{-1}Mpc)\ge 3\times 10^{14}h^{-1}\msun$]
the projection effects in the 0.4-2.4keV and
2.0-10.0keV bands are similar (see Figures 1-10 and Figure 12),
we expect that harder X-ray bands such as that of ASCA
are likely to be more advantageous than softer X-ray bands such as 
that of ROSAT for richer hotter clusters.

Needless to say, all the substructures contain
useful information about the cluster and its surroundings.
Precise comparisons between observations and models
would require applying identical measures 
to both observations and simulations, 
with the latter appropriately including 
large-scale structure, gasdynamics and galaxy formation.

\section{Discussion and Conclusions} 

Utilizing large-scale N-body simulations to investigate
the projection effects on various observables of clusters
of galaxies,
we find that projection alters them in different ways 
to varying extents, but in general projection effects
increase with cluster-centric radius, as expected. 
Even with precise information of the positions of
galaxies in three-dimensional space (sky plane position plus
radial velocity),
quantities constructed from observed positions
and velocities of galaxies in clusters 
suffer from projections of background and foreground structures,  
due to complex motions in and around clusters.
In a rather conservative fashion
we study the X-ray observations of clusters
by assuming
that X-ray clusters are subject only
to the projections of sources within the velocity 
space defined by the cluster galaxies,
and find that contaminations on X-ray properties of clusters 
strongly depend on the observable under consideration.
We summarize the results in six points and conclude 
with a discussion of the limit
of present work and of prospects of significant
future improvement over the current work.

1) The number of galaxies in a cluster is, on average, increased by 
10\% and 20\% at $R_{proj}=(0.5,1.0)h^{-1}$Mpc, respectively, 
due to projection.
The contamination of X-ray cluster luminosity 
in the 0.4-2.4~keV band is 
much smaller, being a 2\% and 8\% increase at the two radii.
The contamination of cluster X-ray 
luminosity at the harder 0.5-10.0~keV band
is comparable to the 0.4-2.4~keV band for clusters with $kT\le 5$keV,
but it is expected
that contamination will be smaller for richer, hotter clusters
($kT > 5$keV) observed with the harder band.
This latter 
finding makes the X-ray cluster luminosity function 
a very simple and useful diagnostic for comparing observations
(Henry \& Arnaud 1991; Henry 1992)
with theoretical predictions
(Kang \etal 1994; Bryan \etal 1994; 
Cen \& Ostriker 1994).

2) For the most massive clusters 
[$M(<1h^{-1}Mpc)\ge 3\times 10^{14}h^{-1}\msun$]
found in the simulations
we find that the 
virial mass estimate (assuming isothermal distribution)
within the radius $R_{proj}\le 1h^{-1}$Mpc,
on average, underestimates the true mass by about 20\%, 
which is in agreement with the mass derived from X-ray temperature
assuming isothermal hydrostatic equilibrium.
The dispersion is somewhat smaller
in the X-ray mass estimate than in the velocity mass estimate.
The gravitational 
lensing reconstructed mass is, on average,
overestimates the true mass by only 5-10\%
but displays a dispersion significantly larger than
that of the X-ray determined mass and comparable to that
of the velocity dispersion determined mass. 
This indicates that cluster X-ray temperature measurements
probably provide a better, stable means 
for galaxy cluster mass determination,
especially when a harder/wider X-ray band (such as ASCA band)
is used where projection contaminations are still smaller.

3) For the richest clusters 
contained in the simulation, 
we show that the ratio of gravitational lensing 
reconstructed mass to velocity (or X-ray temperature) derived mass 
is 1.2-1.3 with a dispersion of $\sim 0.3$ 
within radius $R_{proj}\le 1h^{-1}$Mpc.
Values of $\sim 1.5$ for $M_{lensing}/M_{xray}$ 
are common but values of $\sim 2.0$ are
uncommon for rich clusters
[$M(<1h^{-1}Mpc)\ge 3\times 10^{14}h^{-1}\msun$].
By contrast it is common to have ratios of $2-3$ for poor clusters
[$M(<1h^{-1}Mpc)\le 2\times 10^{14}h^{-1}\msun$].
It seems that the existing disparity of the ratio of
lensing mass to dynamically derived or temperature derived mass
in real observed clusters
can be accounted for by projection effects.
It is, however, never seen in our analysis
that the lensing mass should be
smaller than the velocity or temperature derived mass by
a factor of two.
Note that a velocity bias of galaxies 
over dark matter, observed in some simulations 
(Carlberg, Couchman \& Thomas 1990; Carlberg \& Dubinski 1991; 
Cen \& Ostriker 1992;
Evrard, Summers, \& Davis 1994;
Brainerd \& Villumsen 1994; 
see also Katz, Hernquist \& Weinberg 1992),
not included in the current calculation,
would further raise the 
computed ratio of the lensing to dispersion-based mass. 
Temperature bias, if exists, has the same effect.

4) The gas to total mass ratio in clusters is, on average,
30-40\% higher than the global ratio with a broad distribution
due to projection effects on both quantities involved 
(baryonic mass and total mass) within radius $R_{proj}\le 1h^{-1}$Mpc.
This moderate boost 
narrows the gap but is not sufficient to 
reconcile the standard nucleosynthesis value
of $\Omega_b=0.0125h^{-2}$ (Walker \etal 1991) and $\Omega=1$ 
with the observed gas to mass ratio value in
clusters of galaxies, $0.05h^{-3/2}$, for any reasonable $h$. 
However, it is worth noting that real observations of X-ray clusters,
especially X-ray imaging observations,
may suffer more severe contaminations than we assume here, 
due to our rather conservative assumption
that X-ray contaminations are only limited to the velocity range 
defined by the cluster galaxies.
In any case, it seems 
that the broadness of the observed distribution of the gas to total
mass ratio for real galaxy clusters is adequately explained
by the projection effects alone,
independent of its mean value.

5) We show that substructures in clusters
are significantly affected by projection.
In this particular cosmological model, only about 5\% of clusters
show instrinsic substructure, as measured
by the Dressler-Shectman $\Delta$ statistic,
at $R_{proj}=1.0h^{-1}$Mpc.
But the fraction increases to about 70\% 
when these clusters are ``observed" 
including projection effects.
If this particular cosmological model represents the real universe
in this respect,
the agreement between observed the fraction of real clusters
containing substructure 
and that of our simulated clusters, when subject to projection,
seems to imply that most of the substructures
observed in cluster galaxies are due to projection effects.
X-ray clusters show a similar fraction of clusters with 
intrinsic substructure at $R_{proj}=1.0h^{-1}$Mpc,
if one defines substructure as local X-ray surface brightness
maxima whose luminosity exceeds 10\% of that of the main
(central) structure in the cluster.
In contrast to galaxy maps, X-ray maps are much more
immune to projections in terms of affecting intrinsic substructures
in the clusters,
when an appropriate definition of substructure is chosen.
For example, 
the fraction of clusters that contain no substructure
with X-ray luminosities larger than 10\% of that
of the main structure decreases from 82\% to 78\%
due to projection at $R_{proj}=1.0h^{-1}$Mpc.
We note that 
the contamination in a harder band (such as the ASCA band)
is still smaller than in a softer band.
It therefore seems that X-ray maps provide a better tool
to measure the intrinsic substucture in clusters,
if appropriate measures are devised.
However, measuring the substructure in X-ray maps out to a
large radius such as 
$R_{proj}\ge 2.0h^{-1}$ becomes meaningless,
because projected structures
dominate over the intrinsic substructures.

6) In many cases, galaxy maps and X-ray maps
can be easily misinterpreted because projection of background
or foreground structures may create some illusory situations.
Infall motions of other structures
(which are often not yet bound or not in the process of 
imminent merging onto the 
cluster) towards the main cluster
render it very difficult 
or sometimes impossible to separate them from the intrinsic
cluster structure. 
For example, for the cluster shown in Figure 1,
the binary core structure appearing
as a substructure in the process of
merging onto the cluster at a projected distance of $0.5h^{-1}$Mpc
is due to the projection of a background structure at a
distance of $2-5h^{-1}$Mpc from the main cluster along
the line of sight.
A naive interpretation of 
the existence of such a binary structure would imply
that the main cluster is young, unrelaxed and undergoing 
a major merger, which is obviously incorrect in this case.
Until we can measure real distances of X-ray sources
and galaxies accurately,
we are stuck with the fact that complex motions in the
vicinity of clusters prevent us from fixing the relative
distances between structures along the line of sight.
This fact dictates that
for some observables
the only meaningful way to compare predictions of 
a cosmological model with the cluster observations 
is to subject clusters in a simulated universe 
to exactly the same observational biases and uncertainties,
and to compare the ``observed" simulated clusters
with real ones.

We expect that the projection effects discussed
here depend on the specific parameters of a 
particular cosmological model, namely 
the Hubble constant $h$,
density parameter $\Omega_0$, cosmological constant $\Lambda_0$,
amplitude of the density fluctuations on the cluster
scale $\sigma_8$ and the shape of the power spectrum $P_k$.
While the exact dependence on these parameters
can be made definite only by making more simulations with 
varying input parameters,
we here comment on likely trends.
The results do not directly depend on $h$.
The dependence on the shape of the power spectrum $P_k$ 
is expected to be rather weak within a plausible range.
For a given model (fixing $h$, $\Omega_0$, $\Lambda_0$ and $P_k$)
it is expected that a higher $\sigma_8$ would reduce projection
effects for a cluster above a fixed mass.
For the same reason 
it is expected
that projection effects at high redshift will be
much more severe, which is exacerbated by greater difficulties 
concerning high redshift observations.
The dependence on 
the $\Omega_0$/$\Lambda_0$ combination is more complex
and difficult to predict.
To simplify a bit let us assume that each model reproduces
the observed present-day rich cluster abundance,
and that we are only concerned with CDM models.
Then, we estimate that
for a cluster with a given mass,
the projection effects on quantities which concern only
the total amount of projected matter relative to the main cluster,
such as lensing mass and richness, may be
in the following order from strong to weak at z=0:
$\Omega_0=1$ model, $\Omega_0+\Lambda_0=1$ model,
and open $\Omega_0<1$ model (same $\Omega_0$ as in the $\Lambda$ model).
This conjecture is based on the fact that there is more
matter (in absolute amount) in the intercluster space
in the first model than in the latter two models.
For quantities which depend
on only the baryonic matter in absolute amount (to zero-th order)
such as hot baryonic mass and total X-ray luminosity,
the projection effects may be 
in the following order from strong to weak at z=0:
open $\Omega_0<1$ model,
$\Omega_0+\Lambda_0=1$ model (same $\Omega$ as in the open case),
and $\Omega_0=1$ model.
The above estimate is based on the fact that 
the total amount of 
baryonic matter in each model is the same in physical units 
(e.g., in grams) assuming all models 
obey the standard nucleosynthesis constraint of $\Omega_b$,
but the ratio of baryon to mass is in the indicated order.
For quantities such as substructure the situation is
more complicated because the above two 
factors compete.
In addition, the clusters themselves (without projection effects)
are from smooth to irregular (in the usual sense) in the order:
open $\Omega_0<1$ model,
$\Omega_0+\Lambda_0=1$ model, 
$\Omega_0=1$ model.
Therefore, it is not clear what the net trend will be in this case.
We should await more simulations to make this definitive.

The present study employs an N-body only simulation,
so many of the properties, especially the X-ray properties,
should be taken as crude treatments.
Also, galaxies are picked randomly from dark matter particles, 
although we suspect that the velocity estimate
and mass estimate should
not strongly depend on it.
Hydrodynamic simulations, which incorporate detailed
atomic physics with gasdynamics and gravity 
(Cen \& Ostriker 1992, 1993a,b; 
Katz, Hernquist \& Weinberg 1992;
Babul \& Katz 1993;
Navarro, Frenk \& White 1994;
Summers, Davis, \& Evrard 1995;
Steinmetz \& Muller 1995;
Gnedin 1995;
Katz, Weinberg, \& Hernquist 1996),
should provide us with better tools to study
the effects highlighted here.
One essential requirement for such simulations
is a large simulation box ($L>100h^{-1}$Mpc)
and a sufficiently high resolution 
($\Delta l < 50h^{-1}$kpc) in order 
both to capture the large-scale structure and to
simulate the constituents of clusters.
With rapidly increasing computer power, 
the gradually maturing hydrodynamic cosmological simulation techniques 
and the next generation of galaxy redshift surveys
such as the Sloan Digitial Sky Survey and of X-ray observations
by telescopes such as AXAF
(having both better spatial and spectroscopic resolutions), 
we should not be surprised 
to see leaping advances in our understanding
of clusters of galaxies and their building blocks.

\acknowledgments
The work is supported in part
by grants NAG5-2759, AST91-08103 and ASC93-18185.
The author thanks Frank Summers 
for use of his P$^3$M simulation for present study
and Steven Phelps for a careful reading of the manuscript.
Discussions with Neta Bahcall, George Lake, Bruce Margon, 
Jerry Ostriker and 
Michael Strauss are stimulating and gratefully acknowledged.
The author would like to thank 
George Lake and University of Washington for
the warm hospitality, and financial support
from the NASA HPCC/ESS Program during a visit when
this work was initiated.
Finally, I would like to thank an anonymous referee
for a very constructive and pertinent report.

\newpage
\section*{Figure Captions}
\figcaption[Figure 1]{
shows detailed structure of cluster \#1.
Panel (1) shows the projected distribution of galaxies
in the cluster, as it would be observed after applying the 
pessimistic $3\sigma$ clipping method (Yahil \& Vidal 1977),
whereas Panel (2) shows the ``true" member galaxies of the cluster (see \S 2.2).
Three symbols are used in Panel (1): 
solid dots for ``true" member galaxies [which is also used in Panel (2)
where only ``true" members are shown],
open circles for background galaxies,
and stars for foreground galaxies.
Panel (3) shows the X-ray surface brightness in the band 0.4-2.4keV
due to emission from all sources within 
velocity domain of the cluster, defined by the cluster galaxies.
Panel (4) shows the ``true"
X-ray surface brightness of the cluster due to 
the hot intracluster gas {\it in the cluster only} (i.e., excluding
all possible foreground and background sources).
The contour levels in Panels (3,4) are
$10^{-8,-7,-6, ...}$erg/cm$^2$/sec/sr.
Panels (5) and (6) show the corresponding (emissivity-weighted)
temperature maps for Panels (3) and (4), respectively, 
with contour levels for thick curves of
$10^{7.00,7.25,7.50,7.75, ...}$Kelvin
and thin curves of $10^{6.75,6.50,6.25,6.00, ...}$Kelvin.
Panels (7) and (8) show the galaxies in real space and
in velocity space, respectively.
Panels (9,10) are similar to Panels (1,2)
only for galaxies projected inside the virial radius of the cluster.
Panels (11,12,13,14) are similar 
to Panels (3,4,5,6) but for ASCA $0.5-10.0$keV band.
Panels (15,16) are similar 
to Panels (7,8) but
only for galaxies projected inside the virial radius of the cluster.
Panel (17) shows the galaxy density distribution
in velocity space for ``true" members (solid
histogram) and projected members (dotted histogram) within
$R_{proj}=2.0h^{-1}$Mpc,
and Panel (18) shows the corresponding distributions in real space.
Panel (19) shows 
the line-of-sight velocity as a function of
the radial position for each galaxy  relative to the center
of the cluster
[the symbols have the same meanings as in Panel (1)],
and Panel (20) shows 
the line-of-sight real space position as a function of
the radial position for each galaxy.
\label{fig1}}

\figcaption[Figure 2]{
shows detailed structure of cluster \#2.
See caption of Figure 1 for detailed descriptions.
\label{fig2}}

\figcaption[Figure 3]{
shows detailed structure of cluster \#3.
See caption of Figure 1 for detailed descriptions.
\label{fig3}}

\figcaption[Figure 4]{
shows detailed structure of cluster \#4.
See caption of Figure 1 for detailed descriptions.
\label{fig4}}

\figcaption[Figure 5]{
shows detailed structure of cluster \#5.
See caption of Figure 1 for detailed descriptions.
\label{fig5}}

\figcaption[Figure 6]{
shows detailed structure of cluster \#6.
See caption of Figure 1 for detailed descriptions.
\label{fig6}}

\figcaption[Figure 7]{
shows detailed structure of cluster \#7.
See caption of Figure 1 for detailed descriptions.
\label{fig7}}

\figcaption[Figure 8]{
shows detailed structure of cluster \#8.
See caption of Figure 1 for detailed descriptions.
\label{fig8}}

\figcaption[Figure 9]{
shows detailed structure of cluster \#9.
See caption of Figure 1 for detailed descriptions.
\label{fig9}}

\figcaption[Figure 10]{
shows detailed structure of cluster \#10.
See caption of Figure 1 for detailed descriptions.
\label{fig10}}

\figcaption[Figure 11]{
shows the cumulative
probability distribution of the ratio of 
observed number of galaxies to the true number of galaxies
in the cluster (see \S 2.3),
$n_{proj}/n_{clust}$, for ``clusters" and ``groups" (defined above)
at four projection radii, 
$R_{proj}=(0.5,1.0,2.0)h^{-1}$Mpc and $R_{proj}=R_{200}$.
$R_{200}$ is the radius within which the mean density
of the cluster is 200 times the critical density,
and is individually defined for each cluster.
\label{fig11}}

\figcaption[Figure 12]{
Figure 12a
shows the cumulative
probability distribution of the ratio of 
observed X-ray luminosity to the true cluster X-ray
luminosity due to the hot gas {\it in the cluster},
$L_{x,proj}/L_{x,clust}$ for ``clusters" and ``groups" 
at four projection radii 
$R_{proj}=(0.5,1.0,2.0)h^{-1}$Mpc and $R_{proj}=R_{200}$,
in the 0.4-2.4keV band.
Figure 12b is for the 0.5-10.0keV band.
\label{fig12}}

\figcaption[Figure 13]{
shows the cumulative
probability distribution of the ratio of the 
observed 1-d velocity dispersion to the true 1-d velocity 
dispersion calculated by
considering the true cluster members only, 
$\sigma_{proj}/\sigma_{clust}$,
for ``clusters" and ``groups" 
at four projection radii 
$R_{proj}=(0.5,1.0,2.0)h^{-1}$Mpc and $R_{proj}=R_{200}$.
\label{fig13}}

\figcaption[Figure 14]{
shows the cumulative
probability distribution of the ratio of derived
cluster mass assuming isothermal velocity dispersion (using 
the average observed 
velocity dispersion within the indicated $R_{proj}$)
to the true mass within the indicated radius, 
$M_{VT}/M_{clust}$,
for ``clusters" and ``groups" 
at four projection radii 
$R_{proj}=(0.5,1.0,2.0)h^{-1}$Mpc and $R_{proj}=R_{200}$.
\label{fig14}}

\figcaption[Figure 15]{
shows $M_{VT}/M_{clust}$ as a function of $M_{clust}$ 
for clusters (solid dots) and groups (open circles)
at three different projection radii 
$R_{proj}=(0.5,1.0,2.0)h^{-1}$Mpc.
\label{fig15}}

\figcaption[Figure 16]{
shows $M_{VT}/M_{clust}$ as a function of $n_{proj}/n_{clust}$ 
for clusters (solid dots) and groups (open circles)
at three different projection radii 
$R_{proj}=(0.5,1.0,2.0)h^{-1}$Mpc.
\label{fig16}}

\figcaption[Figure 17]{
shows the cumulative
probability distribution of $M_{xray}/M_{clust}$ 
for clusters [Panel (a)]
and groups [Panel (b)]
at four projection radii 
$R_{proj}=(0.5,1.0,2.0)h^{-1}$Mpc and $R_{proj}=R_{200}$,
where $M_{xray}$ is the mass derived 
from X-ray observation of the luminosity-weighted
temperature of a cluster,
assuming that 
the cluster is in isothermal hydrostatic equilibrium.
\label{fig17}}

\figcaption[Figure 18]{
shows $M_{xray}/M_{clust}$ as a function of $M_{clust}$
for clusters (solid dots) and groups (open circles)
at three different projection radii 
$R_{proj}=(0.5,1.0,2.0)h^{-1}$Mpc.
\label{fig18}}

\figcaption[Figure 19]{
shows the cumulative probability of $M_{lensing}/M_{clust}$, 
where $M_{lensing}$ is mass which gravitational
lensing method should derive (assuming proper calibration 
can be performed), at four projection radii 
$R_{proj}=(0.5,1.0,2.0)h^{-1}$Mpc and $R_{proj}=R_{200}$.
\label{fig19}}

\figcaption[Figure 20]{
shows the cumulative probability of $M_{lensing}/M_{VTc}$, 
where $M_{lensing}$ is mass which gravitational
lensing method should derive,
and $M_{VTc}$ is the virial (isothermal) mass estimate 
using galaxy velocity dispersion,
corrected to its 2-d projected
value (using the individual 3-d density profile of each cluster outside
the indicated radius $R_{proj}$),
for clusters [Panel a()] and groups [Panel (b)]
at four projection radii 
$R_{proj}=(0.5,1.0,2.0)h^{-1}$Mpc and $R_{proj}=R_{200}$.
\label{fig20}}

\figcaption[Figure 21]{
shows $M_{lensing}/M_{VTc}$ as a function $M_{clust}$
for clusters (solid dots) and groups (open circles)
at three projection radii 
$R_{proj}=(0.5,1.0,2.0)h^{-1}$Mpc.
\label{fig21}}

\figcaption[Figure 22]{
shows the cumulative probability of $M_{lensing}/M_{xrayc}$
for clusters [Panel (a)]
and groups [Panel (b)]
at four projection radii 
$R_{proj}=(0.5,1.0,2.0)h^{-1}$Mpc and $R_{proj}=R_{200}$.
where $M_{lensing}$ is mass which gravitational
lensing method should derive,
and $M_{xrayc}$ is the X-ray derived mass
within the 2-d projected radius $R_{proj}$.
\label{fig22}}

\figcaption[Figure 23]{
shows $M_{lensing}/M_{xrayc}$ as a function $M_{clust}$
for clusters (solid dots) and groups (open circles)
at three different projection radii 
$R_{proj}=(0.5,1.0,2.0)h^{-1}$Mpc.
\label{fig23}}

\figcaption[Figure 24]{
shows the differential probability distribution of the
gas to total mass ratio $M_{gas}/M_{tot}$ in units 
of the global mean ratio,
for ``clusters" (thick curve) and ``groups" (thin curve)
at a radius $R_{proj}=1.0h^{-1}$Mpc.
\label{fig24}}

\figcaption[Figure 25]{
shows $M_{gas}/M_{tot}$ as a function of $M_{clust}$
for clusters (solid dots) and groups (open circles)
at three different projection radii 
$R_{proj}=(0.5,1.0,2.0)h^{-1}$Mpc.
\label{fig25}}

\figcaption[Figure 26]{
shows the distribution of the difference between
two kurtoses at a radius $R_{proj}=1.0h^{-1}$Mpc:
$Kurt_{clust}$ is the intrinsic kurtosis
of the cluster velocity distribution without any contamination
and
$Kurt_{proj}$ is the kurtosis of the projected cluster velocity field.
\label{fig26}}

\figcaption[Figure 27]{
plots $Kurt_{proj}$ against $Kurt_{clust}$
for clusters (solid dots) and groups (open circles)
at three different projection radii 
$R_{proj}=(0.5,1.0,2.0)h^{-1}$Mpc.
\label{fig27}}

\figcaption[]{
Figure 28a shows the probability distribution of 
$P(>\Delta)$ for the ``true" cluster without projected external structures,
at three projection radii $R_{proj}=(0.5,1.0,2.0)h^{-1}$Mpc,
where $\Delta$ is the Dressler-Shectman's (1988) 
substructure measure.
The probability is normalized such that the sum of all the bins is unity.
Figure 28b shows the corresponding probability distribution of 
$P(>\Delta)$ for the actual observed cluster including projected members.
\label{fig28}}

\figcaption[]{
Figure 29a shows the distribution of 
$N(L_{sub}>0.1L_{main})$, the number of substructures
each having a luminosity larger than 10\% of that
of the main cluster structure in the 0.4-2.5keV band,
for clusters at three projection radii
$R_{proj}=(0.5,1.0,2.0)h^{-1}$Mpc.
X-ray maps in the band 0.4-2.4keV are used here.
Two curves are shown for the clusters at each projection radius:
the thin curve is generated from the true cluster X-ray maps
due to intrinsic cluster hot gas only;
the thick curve is for 
the corresponding maps including projected structures.
Figure 29b shows the equivalent distribution for groups.
\label{fig29}}

\figcaption[]{
Figure 30a shows the distribution of 
$N(L_{sub}>0.01L_{main})$, the number of substructures
each having a luminosity larger than 1\% (instead of 10\% in Figure 28)
of the main cluster structure in the 0.4-2.5keV band,
for clusters at three projection radii
$R_{proj}=(0.5,1.0,2.0)h^{-1}$Mpc.
Two curves are shown for the clusters at each projection radius:
the thin curve is generated from the true cluster X-ray maps
due to intrinsic cluster hot gas only;
the thick curve is for 
the corresponding maps including projected structures.
Figure 30b shows the equivalent distribution for groups.
\label{fig30}}

\end{document}